\documentclass[runningheads]{llncs}
\usepackage{graphicx}
\usepackage{subfigure}
\usepackage[rightcaption]{sidecap}
\begin{document}
\title{{Automated Segmentation of CT Scans for Normal Pressure Hydrocephalus}
\thanks{This paper was partially supported by the National Institutes of Health [grant number T32-GM08620].}}
%
%
\author{  }
\institute{  }
\author{Angela Zhang\inst{1} \and
Po-yu Kao\inst{1} \and
Ronald Sahyouni\inst{2} \and
Ashutosh Shelat\inst{3} \and
Jefferson Chen\inst{2} \and
B.S. Manjunath\inst{1}}

\authorrunning{A. Zhang et al.}

\institute{University of California, Santa Barbara, Santa Barbara CA 93106, USA 
\and
University of California, Irvine, Irvine CA 93106, USA
 \and
Santa Barbara Cottage Hospital, Santa Barbara CA 93106, USA
\email{angela00@ucsb.edu}
}

\maketitle              
\begin{abstract}
Normal Pressure Hydrocephalus (NPH) is one of the few reversible forms of dementia. Due to their low cost and versatility, Computed Tomography (CT) scans have long been used as an aid to help diagnose intracerebral anomalies such as NPH. However, no well-defined and effective protocol currently exists for the analysis of CT scan-based ventricular, cerebral mass and subarachnoid space volumes in the setting of NPH. The Evan’s ratio, an approximation of the ratio of ventricle to brain volume using only one 2D slice of the scan, has been proposed but is not robust. Instead of manually measuring a 2-dimensional proxy for the ratio of ventricle volume to brain volume, this study proposes an automated method of calculating the brain volumes for better recognition of NPH from a radiological standpoint. The method first aligns the subject CT volume to a common space through an affine transformation, then uses a random forest classifier to mask relevant tissue types. A 3D morphological segmentation method is used to partition the brain volume, which in turn is used to train machine learning methods to classify the subjects into non-NPH vs. NPH based on volumetric information. The proposed algorithm has increased sensitivity compared to the Evan’s ratio thresholding method.

\end{abstract}
\section{Introduction}
NPH presents as ventriculomegaly accompanied with symptoms of dementia, specifically cognitive dysfunction, changes in gait, and urinary incontinence~\cite{Shprecher}. It is estimated that more than 700,000 Americans have NPH. Due to the nonspecific and indolent nature of NPH, the majority of cases are under- or misdiagnosed~\cite{Jaraj}. NPH is one of few reversible causes of dementia in the elderly, making correct diagnosis important, as shunt placement has been demonstrated to be a safe and effective treatment~\cite{Shprecher}.

Current diagnostic methods for NPH involve a mixture of clinical and imaging approaches~\cite{Shprecher}. Although MRI volumetric data may, in ideal cases, provide better detail of the borders of the ventricles than standard CT imaging, shorter acquisition time, use for follow-up, and the sheer number of data points are some advantages of CT imaging in understanding and classifying NPH on a broad scale.

\begin{SCfigure}[5]
\caption{Illustration of Evan’s ratio method. The ratio takes the length of the widest part of the frontal horns (line A) over the length of the widest part of the inner skull (line B). The location of the slice in the z dimension is the location where the parietal lobe appears to be widest.} 
\includegraphics[width=0.15\textwidth]{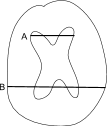}
\label{fig1}
\end{SCfigure}

The current methods available for analyzing brain scans for possible NPH, such as finding the Evan’s ratio, are time-intensive, manual, and prone to error~\cite{Toma}. Evan’s index is the ratio of the transverse diameter of the anterior horns of the lateral ventricles to the greatest internal diameter of the skull in a single slice of a 3D volume. This is illustrated in Figure \ref{fig1}.

Current guidelines state that an Evan’s index of greater than 0.3 indicates NPH. However, recent findings have shown that the Evan’s ratio in fact varies greatly depending on the level (slice location) of the brain CT scan image at which the frontal horns and maximal inner skull diameters are measured~\cite{Toma}. A 3-dimensional, volumetric method of measuring the relevant regions of the brain could help to mitigate these challenges and holds promise for improving NPH differential diagnosis~\cite{Moore}. 

While there are methods to obtain ventricle and cerebral mass volumes in MRI, these methods cannot find the subarachnoid space, as it does not show up in MRI. The method detailed in~\cite{Gunter} automatically detects features of disproportionately enlarged subarachnoid space hydrocephalus in MRI.~\cite{Coupe} uses expert priors to aid in patch based segmentation of the lateral ventricles in MRI. In~\cite{Yepes}, automated ventricular volume measurement in MRI is implemented through the PACS system. The paper claims feasibility in CT, but does not discuss NPH. Another method of lateral ventricle segmentation in MRI is presented in~\cite{Kobashi}. The authors in~\cite{Shao} explore challenges in ventricle segmentation using neural networks in MRI. An automated method of computing the Evan’s ratio from CT is presented in~\cite{Takahashi}, but this method loses the volumetric advantage of directly computing the volumes from CT scans.

This paper proposes a novel method to automatically classify in 3 dimensions the lateral ventricles, cerebral mass, and subarachnoid space from CT scans and use these volumes to predict possible NPH. The proposed method has improved performance compared to the Evan’s Ratio, which can be used as an aid in screening for potential NPH in subjects who might otherwise be misdiagnosed.

\section{Methods and Materials}
\subsection{Data}
The subject data comes from two sources: 
the University of California Irvine Medical Center (UCI) and the Santa Barbara Cottage Hospital. 
This is a retrospective study, with all images de-identified as specified by the IRB agreement between each medical center and 
the University of California, Santa Barbara.

There was no protocol determining the number of slices, orientation, or other imaging parameters for the data used in this study. CT scans of 61 subjects from [Hospital 1 and Hospital 2] 
the University of California, Irvine and Santa Barbara Cottage Hospital
were included in the study, with 34 subjects having a diagnosis of normal and 27 subjects having a diagnosis of NPH. Scans were acquired as part of the treatment process, and the number of slices varied from 25 to 207. For the subjects from 
UCI,
the average subject age is 75 $\pm$ 15 years. For the subjects from 
Cottage Hospital, 
the average age of the subjects is 72 $\pm$ 14 years.

30 manual segmentations were performed by members of the research team under direct supervision and validation by a neurological surgeon. The Evan’s ratio, as measured by or under direct supervision of a neurological surgeon, was measured for all subjects.

\subsection{Algorithm Overview}
\begin{figure}[htp]
\vspace*{-5mm}
\centering
\includegraphics[width=1\textwidth]{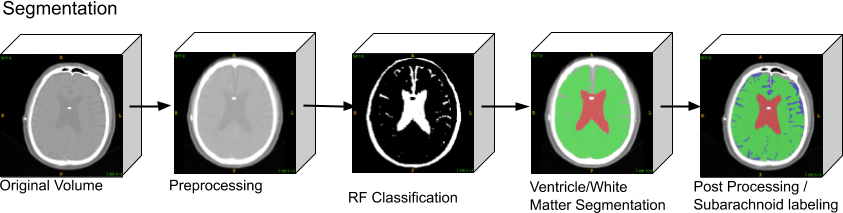}
\caption{Workflow of automatic classification algorithm. From left to right, 1) the original volumetric image. 2) An affine transform is computed to register the volume to a template. 3) RF is used to classify the types of tissue in the volume. 4) Segmentation of ventricle and cerebral mass. 5) Post processing, segmentation of subarachnoid space. Best viewed on screen/in color.} 
\label{fig2}
\end{figure}

There are several major steps involved in the proposed method for NPH prediction, as shown in Figure \ref{fig2}.
The implementation details and code to the algorithm is available at \begin{verbatim}https://github.com/UCSB-VRL/NPH\_Prediction. \end{verbatim}

\subsubsection{Voxel-wise Classification}

\paragraph{Preprocessing}
First, the skull of each scan is extracted using thresholding. The skull volume is used to compute an affine transformation with the standard MRI in MNI152 space~\cite{Jenkinson}. The MNI152 space is a 3-dimensional coordinate system for stereotactic localization in neurosurgery made from the average MRI of 152 scans from the Montreal Neurological Institute of McGill University Health Centre~\cite{Grabner}. The computed affine transform is then applied to its corresponding CT scan. Denoising is applied \cite{Buades}.

\paragraph{Tissue Classification}
The registered scans are used to train a random forest classifier~\cite{Pedregosa}~\cite{Breiman} on a small subset of labeled data ($\sim$10,000 voxels) to recognize the intensity values of 3 different tissue types - cerebrospinal fluid (CSF), cerebral mass, and skull. The classifier then aggregates the votes from the different decision trees to decide the final class of the voxel. There are 4 classes total, including the background class. The trained classifier is then used to classify each voxel of every scan. The trained random forest classifier is used to select relevant regions of the volume and mask the other regions based on tissue type.

\paragraph{Ventricle / Cerebral Mass Segmentation}
Each masked volume of the CSF is then seeded at the center of the anatomical ventricular region based on anatomical prior knowledge of the average ventricle location in MNI152 space. The seeds are then grown using the 3-dimensional Morphological Chan-Vese (MCV) algorithm~\cite{Chan}~\cite{Walt}. This algorithm is a level set evolution algorithm with the goal of minimizing the energy function defined by:
$$F = \mu \times Surface Area(C) + v \times Volume(Inside(C)) + $$

$$\lambda_1 \int_{Inside(C)} |u_0(x,y,z)-c_1|^2 dx dy dz + \lambda_2 \int_{Outside(C)} |u_0(x,y,z)-c_2|^2 dx dy dz,$$ where $u_0(x,y,z)$ is the volumetric image, $\lambda_1, \lambda_2 > 0$ are parameters that can adjust the comparisons between average intensity inside and outside of 3-dimensional surface contour $C$ ($c1$ and $c2$, respectively), and $\mu_1>0$ is a regularizing parameter to promote evolution.

Likewise, each masked volume of the cerebral mass is seeded at three points: on the top, back, and front of the head next to the skull. The boundaries are then found using the MCV algorithm.

Because the ventricular space is separate from the subarachnoid space in terms of fluid flow, this method allows for the separation of the two, even though their tissue classes are the same from the random forest classifier.

\paragraph{Post Processing / Subarachnoid Space Labeling}
Following segmentation of the lateral ventricles and cerebral mass, the remaining voxels (classified as ‘fluid’) are labeled as subarachnoid space. The volumes of each class are computed by multiplying the total number of labeled voxels with the 3D voxel spacing information in the metadata of the scan.

To return the segmentation to the original patient space for purposes of visualisation and segmentation verification, the inverse affine transform from the MNI152 space to the patient space was computed and applied to the automatic segmentation using nearest neighbor interpolation. Each automatic segmentation was then compared to its corresponding manual segmentation in the original patient space. The segmentations performed by the algorithm are compared with 30 manual segmentations of 9 subjects performed by members of the research team under the direct supervision of a neurological surgeon.

\paragraph{Comparison with other methods}
Some basic machine learning methods are implemented to compare with the proposed method. Some of these methods are used as one step in our method. They include random forest classification, 3D morphological geodesic active contours (MGAC), and 3D morphological Chan-Vese (MCV).

The implementation of the alternative methods of ventricle segmentation use thresholding to find the skull region and remove any labels outside of this region. All implementations first compute and apply the affine transformations into MNI152 space, then computes and applies the inverse transformation after completing segmentation. For the morphological chan-vese and morphological geodesic active contour methods, the volumes are seeded in the same manner as the proposed algorithm. The regions are then grown according to their perspective algorithms. Finally, the regions inside the skull not labeled as cerebral mass are then labeled as ventricular space.

\subsubsection{NPH Prediction}

A Support Vector Machine (SVM)~\cite{Cortes} with a radial basis function (RBF) kernel and a Random Forest (RF) classifier are trained and tested on the volumetric information obtained from the segmentation algorithm.

To account for variability in brain size, the total brain volume is calculated by adding the ventricle, cerebral mass, and subarachnoid space volumes. The ventricle, cerebral mass, subarachnoid space, and total brain volumes are used as input features to train a rbf SVM and a random forest classifier. Stratified k-fold cross-validation was used to create training and validation datasets. For each method, 100 classifiers are trained to obtain an average and standard deviation for classifier performance.

For comparison, each scan is also labeled with the Evan’s ratio as measured under direct supervision of a neurological surgeon. NPH prediction on the labeled subset using only the Evan’s ratio are first computed by the current guidelines, with subjects having an Evan’s ratio greater than or equal to 0.3 classified as NPH, and the remaining subjects classified as non-NPH.

\section{Results}
\subsection{Classification Validation and Comparison}
For the scores in Table 1, the Dice Score, $\frac{2|X\cap Y|}{|X|+|Y|} = \frac{2TP}{2TP+FP+FN}$, where $X$ and $Y$ are two classes (positive and negative for each class), is used.

\begin{table}[htp]
\hspace*{-0.5in}
\vspace*{-3mm}
\centering
\caption{Comparison of Dice Scores for various ventricle and cerebral mass segmentation algorithms for CT scans. The scores are reported as mean ± standard deviation.}\label{tab1}
\begin{tabular}{|l|l|l|}
\hline
Method &  Ventricle (Dice) & Cerebral Mass (Dice)\\
\hline
{\bfseries Proposed Method} &  {\bfseries 85.31 $\pm$ 6.16 \%} & {\bfseries 91.03 $\pm$ 2.38 \%}\\
Random Forest &  57.34 $\pm$ 16.52 \% & 88.16 $\pm$ 3.19 \%\\
3D MGAC & 20.51 $\pm$ 19.87 \% & 84.68 $\pm$ 9.13 \%\\
3D MCV & 15.05 $\pm$ 18.36 \% & 85.71 $\pm$ 3.65 \%\\
\hline
\end{tabular}
\end{table}

\begin{figure}
\centering
\includegraphics[width=.4\textwidth]{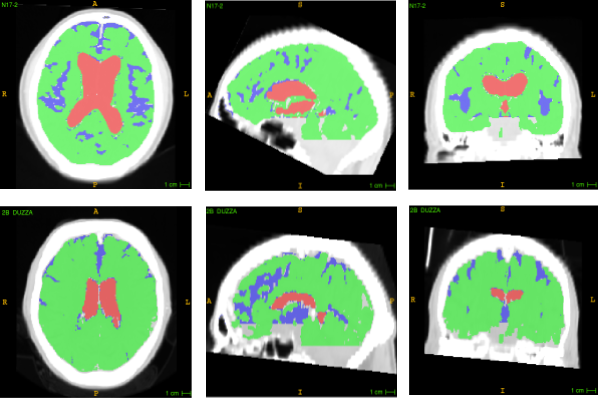}
\caption{Example segmentations generated by our algorithm. The first (top) set of images consists of cross sections of a subject diagnosed with NPH and the second (bottom) set of images are cross sections of a normal subject. Best viewed in color.}
\label{segresults}
\end{figure}

It is important to note that the proposed method is unique in that it allows for separation of ventricle space and subarachnoid space, due to the combination of prior anatomical knowledge, intensity-based tissue classifier, and level set evolution.

https://www.overleaf.com/project/5d003ecae4b5f6444959e88f\begin{figure}
\centering
\includegraphics[width=.65\textwidth]{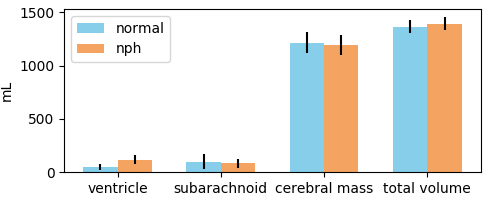}
\caption{Mean and standard deviation of ventricle, subarachnoid, cerebral mass, and total volumes.} \label{barchart}
\end{figure}

Figure \ref{barchart} and Table \ref{tab2} show that the ventricles volumes are greater for NPH, as expected. The subarachnoid space is mostly consistent across subjects. The cerebral mass volumes roughly inversely correspond to ventricle volume. The ventricular volumes of normal subjects are consistent with the average MRI-derived ventricular volumes in the age range of 69.5 $\pm$ 4.8 years~\cite{Jovicich}.

\begin{table}[ht!]
\centering
\caption{Mean and standard deviation of ventricular and cerebral mass volumes.}\label{tab2}
\begin{tabular}{|l|l|l|l|}
\hline
  & Ventricle & Subarachnoid & Cerebral Mass\\
\hline
Normal &  47.4 $\pm$ 28.2 mL & 101.6 $\pm$ 69.7 mL & 1214.6 $\pm$ 100.6 mL \\
NPH & 118.0 $\pm$ 41.2 mL & 85.2 $\pm$ 44.3 mL & 1210.2 $\pm$ 95.6 mL\\
\hline
\end{tabular}
\end{table}

\begin{SCfigure}[2]
\centering
\includegraphics[width=.52\textwidth]{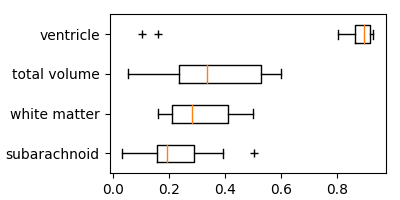}
\caption{Box-and-whisker plots showing the significance of each feature for the RBF SVM model. It is evident that the ventricular volume is the most important feature, which follows expectation.} \label{features}
\end{SCfigure}

\subsection{Diagnosis Scores and Comparison with Evan’s Ratio}

\begin{table}[ht!]
\centering
\caption{NPH prediction scores using all volume information for SVM and RF classifiers compared with Evan's Index thresholding. The train/test split is 50/11.}\label{tab3}
\begin{tabular}{|l|l|l|}
\hline
  & Sensitivity (Train/Test) & Specificity (Train/Test)\\
\hline
Evan’s Index, Thresholding &  75\% (all data) & \textbf{89\% (all data)} \\
\textbf{Vol. Features, RBF SVM} & \textbf{90 $\pm$ 5 / 86 $\pm$ 13\%} & 89 $\pm$ 3 / 85 $\pm$ 10\%\\
Vol. Features, RF & 99 $\pm$ 2 / 86 $\pm$ 14\% & 96 $\pm$ 2 / 84 $\pm$ 10\%\\
\hline
\end{tabular}
\end{table}
The SVM with RBF kernel used parameters C=2, gamma=0.1. The RF classifier used 200 estimators, had a minimum sample split of 3, gini criterion, max depth of 4, and max features of 2.

As seen in Table \ref{tab3}, the SVM had the best performance in NPH sensitivity, but slightly lower specificity than Evan's index. From Figure \ref{features}, all of the volumetric information was used by the SVM to predict possible NPH.

\section{Discussion and Future Work}
The paper presents a fully automated, volumetric method of lateral ventricles, subarachnoid space and cerebral mass segmentation in CT scans. Additionally, this paper proposes a fully automated, volumetric method to predict NPH diagnosis, which in conjunction with the clinical symptomatology, can facilitate the diagnosis of NPH and rule-out subjects who do not meet the radiographic criteria of an NPH diagnosis. This technological system outperforms the thresholding method using Evan’s ratio and can be used as a screening tool to identify or stratify possible NPH cases in a clinical setting.

The work presented in this paper is intended as a proof of concept, with representative samples of CT scans from subjects in each category. In order to achieve more robust classification and higher model scores, we plan to collect a greater number of samples over time.

Additionally, it may be beneficial to explore the combination of MRI and CT scans to create a multimodal method for more fine-tuned diagnosis as well as for symptoms and treatment outcome prediction. This algorithm could be further refined by incorporating relevant demographic and medical variables.

%
%
%
%
\begin{footnotesize}

\end{footnotesize}
\end{document}